\documentclass[draft,jgrga]{agutex}
\usepackage{lineno,amsfonts,amsmath,graphicx,color,ulem, upgreek}

 \usepackage{lineno}

   \setkeys{Gin}{draft=false}
\authorrunninghead{MEDVEDEV ET AL.}

\titlerunninghead{MARS THERMOSPHERE COOLING}

\authoraddr{Corresponding author: A. S. Medvedev,
  Max Planck Institute for Solar System Research,
  Justus-von-Liebig-Weg 3, 37077 G\"ottingen, Germany.
  (medvedev@mps.mpg.de)}

\begin{document}
\title{Cooling of the Martian thermosphere by CO$_2$ radiation and gravity
  waves: An intercomparison study with two general circulation models}



\authors{Alexander S. Medvedev,\altaffilmark{1,2}
Francisco Gonz\'{a}lez-Galindo,\altaffilmark{3}
Erdal Yi\u{g}it,\altaffilmark{4}
Artem G. Feofilov,\altaffilmark{5}
Francois Forget,\altaffilmark{5}
and Paul Hartogh\altaffilmark{1} }

\altaffiltext{1}{Max Planck Institute for Solar System Research, G\"ottingen,
 Germany}

\altaffiltext{2}{Institute of Astrophysics, Georg-August University,
G\"ottingen, Germany}

\altaffiltext{3}{Instituto de Astrofisica de Andaluc\'ia, CSIC, Granada,
Spain}

\altaffiltext{4}{Space Weather Group, School of Physics Astronomy and
 Computational Sciences, Fairfax, VA, USA}

\altaffiltext{5}{Laboratoire de M\'et\'eorologie Dynamique, CNRS, Paris,
 France.}


\begin{abstract}
  Observations show that the lower thermosphere of Mars ($\sim$100--140 km)
  is up to 40 K colder than the current general circulation models (GCMs) 
  can reproduce. Possible candidates for physical processes missing in the 
  models are larger abundances of atomic oxygen facilitating stronger CO$_2$ 
  radiative cooling, and thermal effects of gravity waves. Using two 
  state-of-the-art Martian GCMs, the Laboratoire de M\'et\'eorologie Dynamique 
  and Max Planck Institute models that self-consistently cover the atmosphere 
  from the surface to the thermosphere, these physical mechanisms are 
  investigated.  Simulations demonstrate that the CO$_2$ radiative cooling 
  with a sufficiently large atomic oxygen abundance, and the gravity 
  wave-induced cooling can alone result in up to 40 K colder temperature in 
  the lower thermosphere. Accounting for both mechanisms produce stronger 
  cooling at high latitudes. However, radiative cooling effects peak above 
  the mesopause, while gravity wave cooling rates continuously increase with 
  height. Although both mechanisms act simultaneously, these peculiarities 
  could help to further quantify their relative contributions from future 
  observations.
\end{abstract}



\begin{article}


\section{Introduction}
\label{sec:intro}

The lower thermosphere of Mars ($\sim$100--140 km) is the gateway between the
atmosphere and space. It is affected simultaneously from below and above by 
various dynamical and radiative processes. Spacecraft decelerate and modify 
their orbits in this region by dipping into denser layers. The knowledge of
processes that control the lower thermosphere, and the ability to predict its
state and evolution are of great importance for planning and performing future
aerobraking operations. Temperatures and densities in this region also regulate 
the physical processes responsible for atmospheric escape \citep{vale09,yagi12}.
A precise characterization of the lower thermosphere is thus a necessary step 
towards a better understanding of the long-term evolution of the Martian 
atmosphere.

Recent observations have shown that the lower thermosphere is systematically
colder than general circulation models (GCMs) predict \citep{forg09,mcdunn10}. 
This discrepancy indicates that there is a major gap in our understanding of 
the physics and dynamics of the lower thermosphere, which is addressed in this 
paper.

The most comprehensive dataset to date covering the lower thermosphere has been 
collected with the SPICAM (Spectroscopy for the Investigation of the
Characteristics of the Atmosphere of Mars) instrument onboard the Mars Express
(MEX) orbiter \citep{bert06,forg09}. It contains vertical profiles of
temperature and density at altitudes between 50 and 140 km retrieved from
observations of stellar occultations in UV \citep{quem06}. A comparison with
simulations using the Laboratoire de M\'et\'eorologie Dynamique (LMD) Mars GCM
has revealed that the observed temperatures are, generally, 10 to 40 K lower
than those simulated, and that this difference apparently increases with 
height \citep{forg09}. Further intercomparison with the MEX/SPICAM data has been
performed by \citet{mcdunn10} using an alternative modeling framework -- the
Mars Thermosphere General Circulation Model (MTGCM) coupled with the NASA Ames
Mars General Circulation Model (MGCM). They also found that the simulated
atmosphere above the mesopause was consistently warmer, and that the model
mesopause was often too low. Accelerometers onboard spacecraft during
aerobraking phases are an additional, although limited, source of data on
density and temperature in the lower thermosphere \citep{Keating98,tolson02,
tolson07,withers06}. These measurements have been compared with the
predictions from MTGCM--MGCM \citep{bough06} and LMD--MGCM \citep{gonz09b}.
The two models agreed well in simulating the latitudinal cross-sections of
temperature at $\sim$120 km, but, again, they turned out to be approximately
35 K higher than those observed \citep[Figure 9]{gonz10}.

Certainly, such warm model biases require an explanation. One of the offered
reasons links the errors to the uncertainties in specifying the airborne dust.
\citet{bell07} have shown that simulations in the mesosphere and lower
thermosphere are very sensitive to distributions of aerosol below. Adjusting
the amount of the airborne particles and/or their physical properties can
nudge simulations closer to observations during certain periods and locations
\citep{forg09,mcdunn10}. To estimate global effects of atmospheric dust,
\citet{mykh13} performed simulations with the Max Planck Institute (MPI) GCM
using the aerosol optical depth measured by the Mars Climate Sounder onboard
Mars Reconnaissance Orbiter (MCS--MRO), and by the Thermal Emission
Spectrometer onboard Mars Global Surveyor (TES--MGS). They found that, during
major dust storms, temperature in the lower thermosphere drops by several tens
of Kelvin degrees, which is comparable to the systematic model bias mentioned
above. However, there are two major shortcomings with using only the diabatic 
effects of lower atmospheric dust to explain the systematically warmer 
mesosphere simulated by GCMS.  First, the dust-induced cooling in the 
mesosphere and lower thermosphere (MLT) occurs only during storms. Second, not 
all dust storms result in a uniform cooling of the lower thermosphere. 
The response is more complex: polar regions during equinoctial storms, and low 
latitudes during solstitial events, on the contrary, experience warmings.

Clearly, key physical mechanisms are missing in current GCMs. Two candidates 
have emerged to date to reconcile the simulated temperatures with the 
observations by SPICAM in the lower thermosphere: the CO$_2$ radiative cooling, 
and gravity wave (GW) heating/cooling. In this paper, we analyze them with two 
currently available Martian GCMs, which self-consistently cover the atmosphere 
from the surface to the thermosphere: the LMD and MPI MGCMs. They are based on 
different dynamical cores, and employ largely independent suites of physical
parameterizations. These differences allow us to separate the effects imposed 
by each mechanism from possible deficiencies of individual GCMs.

The outline of this paper is the following. Section~\ref{sec:back} presents 
the scientific background and past work concerning the role of CO$_2$ cooling 
and GW heating/cooling in the atmosphere of Mars. The numerical tools (MPI, 
LMD GCMs, and the GW scheme) used in this study are briefly described in 
Section~\ref{sec:GCMs}. The setup of the models and numerical experiments are 
outlined in Section~\ref{sec:setup}. Results of simulations with horizontally 
uniform (one-dimensional) vertical profiles of atomic oxygen volume mixing 
ratio are described in Section~\ref{sec:1D}, while effects of latitudinal 
variations of oxygen are analysed in Section~\ref{sec:2D}. Net heating/cooling
rates and their relationship with the meridional circulation are discussed in 
Section~\ref{sec:net}. The simulated neutral densities are presented in 
Section~\ref{sec:dens}, and the conclusions are drawn in Section~\ref{sec:conc}.

\section{Background and Past Work}
\label{sec:back}

Atmospheric cooling by CO$_2$ is created by radiative transfer in the 
15-$\mu$m band. Under the breakdown of local thermodynamic equilibrium 
(non-LTE), collisions with atomic oxygen O facilitate energy exchange between 
the kinetic energy of CO$_2$ molecules (temperature) and their excited 
vibrational states. Hence, cooling rates are proportional to [O] abundances 
and the CO$_2$-O quenching rate coefficient, $k_{VT}$. Observations of [O] in 
the Martian atmosphere are sparse, and those available show a substantial 
variability \citep{stewart92}. The problem is aggravated by the uncertainty of 
the $k_{VT}$, the value of which differs by a factor of 3--4 between 
laboratory measurements and atmospheric estimates on Earth 
\citep{feofilov2012}. At the moment, both MPI and LMD models use the ``median" 
value of $k_{VT}=3.0 \cdot 10^{-12}$ cm$^3$~s$^{-1}$, but the exact value is 
an open and important question for all CO$_2$-containing planetary atmospheres.
These uncertainties with the GCM input, especially in the atomic oxygen density 
that varies by orders of magnitude, translate into the ambiguity with cooling 
rates, and, ultimately, with simulated temperatures.
According to this hypothesis, the overestimation of temperature in the lower
thermosphere is due to the underestimation of [O] abundances \citep{forg09}.
As described in section \ref{sec:1D}, the parameterizations that simulate the
CO$_2$ cooling in Martian GCMs have traditionally used a prescribed constant
atomic oxygen profile. \citet{forg09} showed how using instead the temporally
and spatially varying atomic oxygen abundance predicted by the LMD--MGCM
produced significantly lower temperatures in the lower thermosphere, which
were in a better agreement with SPICAM measurements. \citet{mcdunn10} have
also illustrated with the MTGCM--MGCM framework that an increase of atomic
oxygen volume mixing ratio by 50\% yielded 5 to 10 K lower model temperatures.

The other mechanism pertains to cooling by internal gravity waves (GWs) that 
are continuously generated by a variety of meteorological processes in the
lower atmosphere, such as, flow over topography, convection, storms, and front
systems. GWs effectively propagate upward transporting momentum and energy,
and their amplitudes exponentially grow with height, compensating for the 
decay of density. At higher altitudes, they become increasingly unstable, and 
break and/or dissipate due to an intensification of dissipative processes, 
such as, nonlinear interactions, molecular diffusion and thermal conduction
\citep[][and reference therein]{YigitMedvedev15}. Thus, gravity wave momentum 
deposition to the mean flow, or ``gravity wave drag", causes wind acceleration 
or deceleration. Mechanical energy of obliterating waves irreversibly converts 
into heat, and, thus, affects temperature. In addition, dissipating GWs induce 
a downward sensible heat flux. In a freely propagating GW harmonic, 
oscillations of temperature ($T^\prime$) and vertical velocity ($w^\prime$) 
are in opposite phase, and the mean sensible heat flux associated with the 
harmonic $\overline{w^\prime T^\prime}=0$. Dissipation introduces a phase 
shift between $T^\prime$ and $w^\prime$, and $\overline{w^\prime T^\prime}$ is 
no longer zero, always negative, that is, directed downward. Vertical 
divergence of the wave-induced sensible heat flux, $-\rho^{-1}
d(\rho\overline{w^\prime T^\prime})/dz$, enters the thermodynamic equation for
the mean temperature, and represents the cooling/heating rate: cooling
above the dissipation level, and heating below \citep{MK03,YigitMedvedev09,
MY12}. Such waves have horizontal scales from tens to hundreds of kilometers, 
which, in most cases, are not resolved by GCMs. Therefore, the effects of 
subgrid-scale GWs in GCMs must be parameterized.

The most recent and comprehensive parameterization originally developed for 
and utilized in Earth GCMs extending from the lower atmosphere into the 
thermosphere is the spectral nonlinear GW scheme of \citet{Yigit_etal08}. 
\citet{myh11} have successfully applied this scheme to estimate GW effects in 
the Martian thermosphere. These effects were shown to be significant, and can 
thus not be neglected above $\sim$100 km. 
Further interactive simulations with the \citet{Yigit_etal08} scheme 
implemented into the MPI--MGCM have demonstrated that GWs dramatically change 
the lower thermospheric circulation. Their dynamical effects are reminiscent 
of those in the terrestrial MLT \citep{myhb11}. The MGCM simulations of 
\citet{MY12} have also shown that GW-induced thermal effects yielded the
necessary cooling rates in order to reproduce the lower thermospheric
temperatures that are 10--40 K colder in observations than in previous model
simulations. Therefore, their simulations closely match the SPICAM measurements
analyzed in the work of \citet{forg09}. The inclusion of parameterized effects
of GWs into the MPI--MGCM has allowed it to reproduce for the first time the
latitudinal night-time temperature structure at $\sim$120 km derived from
aerobraking measurements of \citet{bough06}. These simulations predicted, in
particular, that most of the cooling takes place at high latitudes, which 
cannot be validated at the moment due to an insufficient coverage of the 
SPICAM database.

\section{Numerical Tools}
\label{sec:GCMs}

\subsection{The Max Planck Institute General Circulation Model}
\label{sec:mpigcm}

This GCM is based on a spectral dynamical core, and employs physical
parameterizations described in detail in the works of \citet{hart05,hart07},
and \citet{MH07}. The vertical domain extends from the surface to
approximately 150--160 km (model top pressure $p=3.6\times 10^{-6}$ Pa).
It is represented by 67 hybrid levels: terrain-following near the surface,
and pressure-based in the middle and upper atmosphere. The simulations to be
presented here have been performed at a T21 horizontal spectral truncation,
which is equivalent to a 64 $\times$ 32--gridpoint resolution in longitude
and latitude, correspondingly.

Heating and cooling rates due to the radiative transfer in the gaseous
CO$_2$ are calculated with separate parameterizations for LTE and non-LTE
conditions. In the lower atmosphere, an LTE radiation scheme of
\citet{nakaj00} based on the $k$-distribution method is used. In the middle
and upper atmosphere, the exact non-LTE code ALI-ARMS of \citet{kutepov98,
gusev03} optimized with respect to a number of vibrational levels involved in 
the problem and with the opacity distribution function method  
\citep{feofilov2006,feofilovkutepov2012} is applied for calculations in the
CO$_2$ 15-$\mu$m band. The accuracy of the optimized model for cooling/heating
rate calculations above 70~km altitude is $\approx$10--15\% with respect to 
line-by-line calculations. The profiles of heating/cooling rates from the LTE 
and non-LTE schemes are smoothly merged between 60 and 70 km. Heating rates 
due to absorption of solar radiation in the near-IR bands of CO$_2$ are 
accounted for with a simple parameterization given by formulae 1 and 2 of 
\citet{gonz09}. Heating due to absorption of UV and EUV solar radiation by 
CO$_2$ in the upper atmosphere is calculated for 37 spectral intervals between 
5 and 105 nm. The scheme uses the solar EUV flux model of \citet{richards94}, 
and the heating efficiency 0.22 \citep{fox96}.

The model employs the MSTRN-X radiative scheme \citep{nakaj86} for
calculations of heating and cooling rates due to absorption, scattering and
emission by atmospheric dust. The scheme uses 19 representative wavelength
bands: 9 in the visible and 10 in the IR spectral range. The adopted optical
properties of dust particles are described in the work of 
\citet[section 3.3]{hart05}.

\subsection{The Laboratoire de M\'et\'eorologie Dynamique General Circulation
Model}
\label{sec:lmdgcm}

The LMD--MGCM solves the primitive equations using a grid-point discretization
by a dynamical solver inherited from the terrestrial LMDZ GCM \citep{forg99}.
It is a ground-to-exosphere (top at about 250 km from the surface) model. The
grid used for this work includes 64 $\times$ 48 (in longitude and latitude)
points in the horizontal, with 49 vertical hybrid levels.

The physical parameterizations included in the LMD--MGCM are described in
\citep{forg99,mont04,lefe04}, and for processes important in the upper
atmosphere in \citep{ange05,gonz05,gonz09}. The latest model improvements
that concern the upper atmosphere and described by \citet{gonz13}, have also
been included in the simulations presented here. In particular, the improved
15-$\mu$m cooling scheme using five CO$_2$ bands, a calculation of the full
exchange between atmospheric layers, and the possibility of including the
spatially and temporally variable atomic oxygen density as self-consistently 
provided by the photochemical model described in \citep{gonz09,gonz13}, is used 
in this work. Note, however, that for the simulations included here, constant 
atomic oxygen profiles as described in sections \ref{sec:1D} and \ref{sec:2D} 
are used instead.
For the UV heating, the scheme described in \citet{gonz05} is used with a 
heating efficiency of 0.22, identical to that employed in the MPI--MGCM.

The LMD--MGCM includes a parameterization of effects of subgrid-scale
orographic gravity waves, as described in in the work of \citep{forg99}. 
Previous works \citep{ange05}, however, have shown that the effects of such 
waves in the Martian thermosphere are negligible. A parameterization for 
non-orographic gravity waves is not included in the current version of the 
model.

\subsection{The Extended Gravity Wave Parameterization}

This GW parameterization has been described in full detail in the work by
\citet{Yigit_etal08}, and its implementation to the MPI--MGCM and setup are
given in the work of \citet{myhb11}. The scheme solves the equation for the
vertical propagation of horizontal momentum fluxes $F$ for subgrid-scale GW
harmonics, and accounts for their refraction by larger-scale wind and
temperature, dissipation and breaking. For a single harmonic $j$ of the
spectrum, the equation has the form
\begin{linenomath*}
\begin{equation}
\frac{dF_j}{dz}= -\biggl( \frac{\rho_z}{\rho} + \beta^j_{non} + \beta^j_{mol}
      \biggl) F_j,
\label{eq:gwflx}
\end{equation}
\end{linenomath*}
where $F_j=\overline{u^\prime_j w^\prime_j}$ is the momentum flux per unit mass;
$u^\prime$ and $w^\prime$ are the wave perturbations of the horizontal and
vertical winds, correspondingly; the overline denotes an appropriate averaging
over subgrid scales; $\rho$ is the altitude dependent neutral mass density,
where the subscript ``z" denotes a derivative with respect to the altitude. The
vertical damping rates $\beta^j$ for a given harmonic in (\ref{eq:gwflx}) are
due to saturation/breaking associated with nonlinear effects in the individual
harmonic, and/or caused by interactions with others in the spectrum
($\beta^j_{non}$), and due to molecular diffusion and thermal conduction
($\beta^j_{mol}$). The vertical profiles of $F_j$ are formed by a competition
between growing wave amplitude in response to the decreasing density, and the
decay due to dissipation and/or breaking. The momentum flux divergence for a
particular harmonic yields the momentum deposition to the mean flow, $a_j =
\rho^{-1} d(\rho F_j)/dz$, and the net acceleration/deceleration consists of
contributions of all $a_j$.  Similarly, dissipating GW harmonics produce 
heating/cooling rates, which affect the mean temperature, that is, the energy 
balance of the neutrals. They can conveniently be expressed in terms of $a_j$
\citep[][Equation 36]{MK03}:
\begin{linenomath*}
  \begin{equation}
    E^j=c_p^{-1}a_j(c_j-\bar{u}), \quad \quad
    Q^j=\frac{H}{2R\rho}\frac{d}{dz} \bigl[\rho(c_j-\bar{u})\bigr],
    \label{eq:gw-heat}
  \end{equation}
\end{linenomath*}
where $E^j$ is the irreversible heating due to conversion of the wave
mechanical energy into heat, $Q^j$ is the differential cooling/heating due to
the wave-induced sensible heat flux, $c_p$ is the specific heat at constant
pressure, $H$ is the density scale height, $\bar{u}$ is the mean
(GCM-resolved) wind, and $c_j$ is the horizontal phase speed of the $j$-th
harmonic. The total heating and/or cooling rate is the sum of contributions
of individual harmonics. Within the framework of the GW parameterization,
Equations (\ref{eq:gwflx}) and (\ref{eq:gw-heat}) are solved interactively,
and the calculated tendencies are accounted for in the momentum and
thermodynamic energy equations of the GCM.

Solution of (\ref{eq:gwflx}) requires a specification of $F_j$ at a certain
height. Since most of GWs are being generated in the lower atmosphere, we
prescribed the source level at $p=260$ Pa. The incident spectrum in the
simulations was represented by 28 harmonics, whose horizontal phase speeds were
distributed normally around the local wind speed. The magnitudes of the fluxes
were normalized to match the observed background GW-related variances on Mars
\citep{creasey06}. Such choice of the source spectrum is based on the extensive
experience with applications of the extended GW scheme to GCM studies on Earth
\citep[e.g.,][]{Yigit_etal12, Yigit_etal14, YigitMedvedev09, YigitMedvedev12}
and Mars \citep{myhb11,mykh13,MY12}, and is discussed in detail there.

\section{Model and Experiment Setup}
\label{sec:setup}

In order to exclude a possible influence from below associated with dust
variability, we performed simulations for low-dust equinox period around
$L_s=0^\circ$. The aerosol was taken to be well mixed vertically, and
prescribed by the \citet{conr75} formula:
\begin{linenomath*}
\begin{equation}
q(z)=q_0 \exp \Biggl\{ \nu \Biggl[ 1- \biggl(\frac{p_0}{p}
      \biggr) \Biggr] \Biggr\},  \qquad\qquad
    p<p_0,
    \label{eq:conr}
  \end{equation}
\end{linenomath*}
where $q$ is the dust mixing ratio, $q_0=q(z=0)$, $p$ is pressure, $p_0=610$
Pa is the global-mean surface pressure, and the Conrath parameter $\nu=0.007$
corresponds to a practical absence of dust radiative effects above 60--70 km.
$q$ was normalized such that the total optical depth in visible wavelengths
$\tau$ was uniformly equal to 0.2. Simulations were conducted for the solar
activity close to a minimum (solar flux index $F_{10.7}= 80\times 10^{-22}$
W~m$^{2}$~Hz$^{-1}$).

In each scenario described below, the GCMs were run for at least 30 sols
preceding $L_s=0^\circ$ to exclude adjustment processes, and the outputs from
15 sols immediately following this moment were analyzed. Therefore, unless
stated otherwise, the figures are based on 15-sol averaged fields.

\section{Results with One-Dimensional Oxygen Profiles}
\label{sec:1D}

\begin{figure}
\centering
\includegraphics[width=0.55\columnwidth,clip]{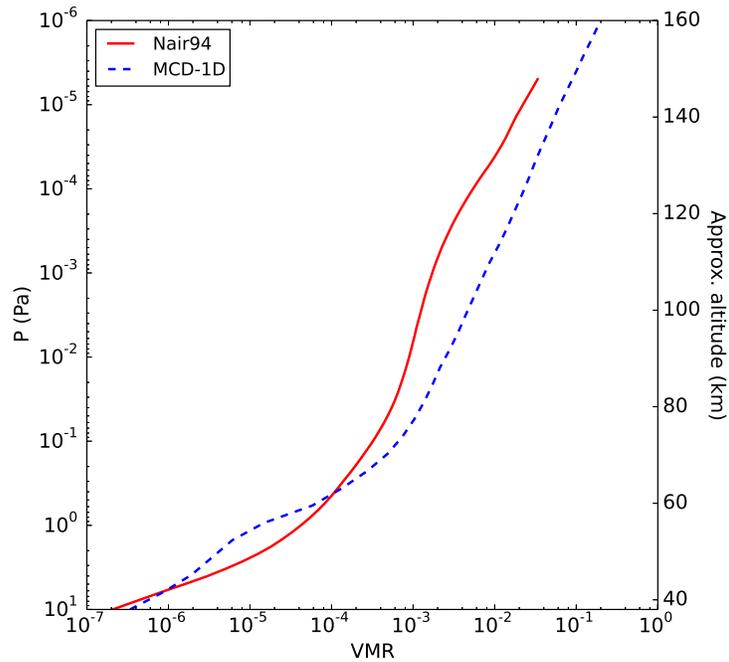}
\caption{Profiles of atomic oxygen volume mixing ratios (in ppm) for the
``low O" \citep{Nair94} (red solid line), and ``enhanced O" MCD-1D (blue
dashed line) scenarios. Geometric heights of pressure levels are highly 
variable, but an approximate (spatially and temporally averaged) 
correspondence between them is plotted here, and given in Table~\ref{tab:1}. 
(1 Pa = 10 microbar).}
\label{fig:1DOx}
\end{figure}

The only \textit{in situ} measurements of atomic oxygen in the Martian
thermosphere have been performed with the neutral mass spectrometers during
the descents of Viking 1 and 2 in the altitude range of 120 km 
\citep{hanson77}. Other constraints on atomic oxygen abundances have been 
inferred from UV airglow observations \citep[e.~g.,][]{stewart92,huestis08}. 
The measurements as well as photochemical modeling have demonstrated that 
atomic O concentrations are highly variable, and depend on local time, 
geographical location, season, solar activity, etc. In this section, we 
explore the response of the atmosphere to gross changes of volume mixing 
ratios [O] represented by two horizontally uniform and temporally constant 
vertical profiles, which are plotted in Figure~\ref{fig:1DOx}. The first one 
(shown in red) is taken from photochemical simulations of \citet{Nair94}. 
It was routinely adopted for
simulations with many Martian GCMs utilizing the non-LTE CO$_2$ cooling
parameterization of \citet{lval01} (in the so-called ``static oxygen" version)
\citep[e.g.,][]{ange05,bell07,gonz10,mcdunn10}. A very similar distribution of
[O] was used in simulations with the MPI--GCM, although in conjunction with the
different radiation scheme \citep[e.g.,][]{hart05,MH07,myhb11,mykh13}. The
second profile (plotted with blue dashed lines, and hereafter referred to as
MCD-1D) has been obtained by averaging the output from the Mars Climate
Database (MCD). The latter is based on the LMD--MGCM simulations with
interactive photochemistry, and provides distributions of [O] as functions of
latitude, longitude, pressure level, and local time for each Martian ``month",
for $L_s=0-30^\circ$ in our particular case. The MCD-1D profile describes a
scenario with an increased atomic oxygen abundance -- the volume mixing ratio
is greater than that of \citet{Nair94} everywhere in the middle and upper
atmosphere: by about a factor of 8 at $p=10^{-3}$ Pa, and $\sim$4 at
$p=10^{-5}$ Pa. Despite such significant differences, both profiles are well
within the limits of variations that follow from photochemistry models, and
from a limited number of observations.

\subsection{Impact of Atomic Oxygen Variations}

\begin{figure}
\centering
\includegraphics[width=1.0\columnwidth,clip]{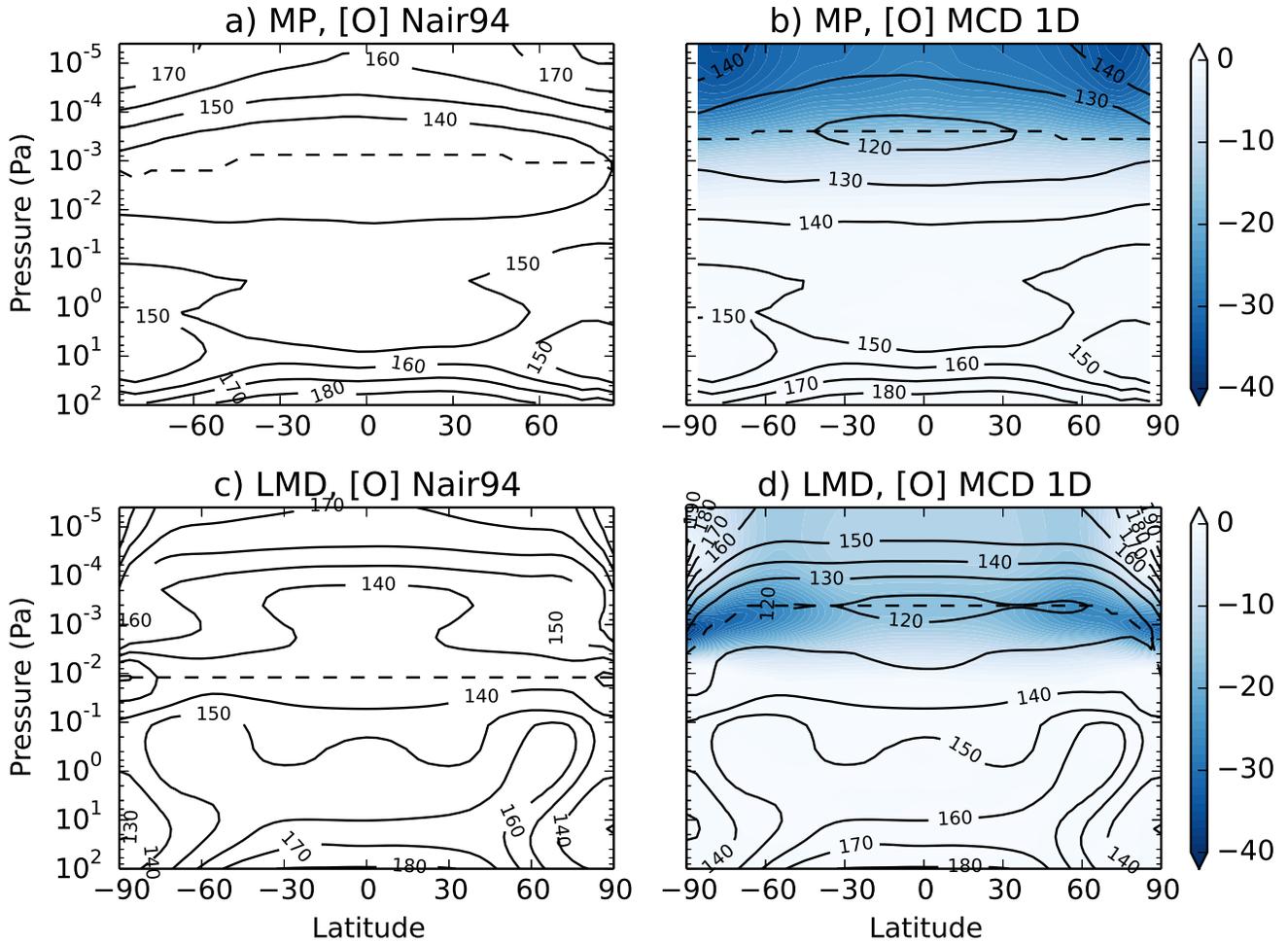}
\caption{Zonal-mean temperature (contours) simulated without inclusion of
gravity wave effects by the MPI--GCM (upper row) and LMD--MGCM (lower row).
Results for the ``low" oxygen \citep{Nair94} scenario are shown in the left
column, and for the MCD-1D run are in the right column. Color shading denotes
the temperature difference between the latter and the former scenarios.
Black dashed line indicates the mesopause.}
\label{fig:1D-old}
\end{figure}

To compare the net effects on temperature introduced by these profiles, the
first series of simulations have been performed without including the GW
parameterization. The corresponding zonal mean temperatures are presented
in Figure~\ref{fig:1D-old}. In the runs with the low oxygen abundance (marked
as ``Nair94"), the simulated temperatures differ by not more than 15 K, except 
in the vicinity of the poles (Figure~\ref{fig:1D-old}a,c). Both models 
simulated a relatively ``warm" mesopause, and placed it between
$p=10^{-2}$ and $10^{-3}$ Pa. The two simulations also yield very close 
temperature
gradients in the lower thermosphere, especially in low and middle latitudes.
When the runs were repeated with the MCD-1D atomic oxygen profile (``large"
O abundances), the simulated temperatures expectedly dropped in the upper
portions of the domains. The differences with the ``low oxygen" runs are
highlighted with blue shades in Figure~\ref{fig:1D-old}b,d. It is seen that
the differences are the largest in middle and high latitudes ($\sim -40$ K),
and are around --20 K elsewhere in the lower thermosphere.
Accordingly, the mesopauses in the MCD-1D simulations are about 20 K colder
(120 K), and located somewhat higher (at approximately $10^{-3}$ Pa).
There are also differences in the response of both models to the increased [O].
It is by $\approx$15 K greater above the mesopause in the MPI model, and
by about same amount larger at the mesopause in the LMD--MGCM.

\begin{figure}
\centering
\includegraphics[width=0.65\columnwidth,angle=-90,clip]{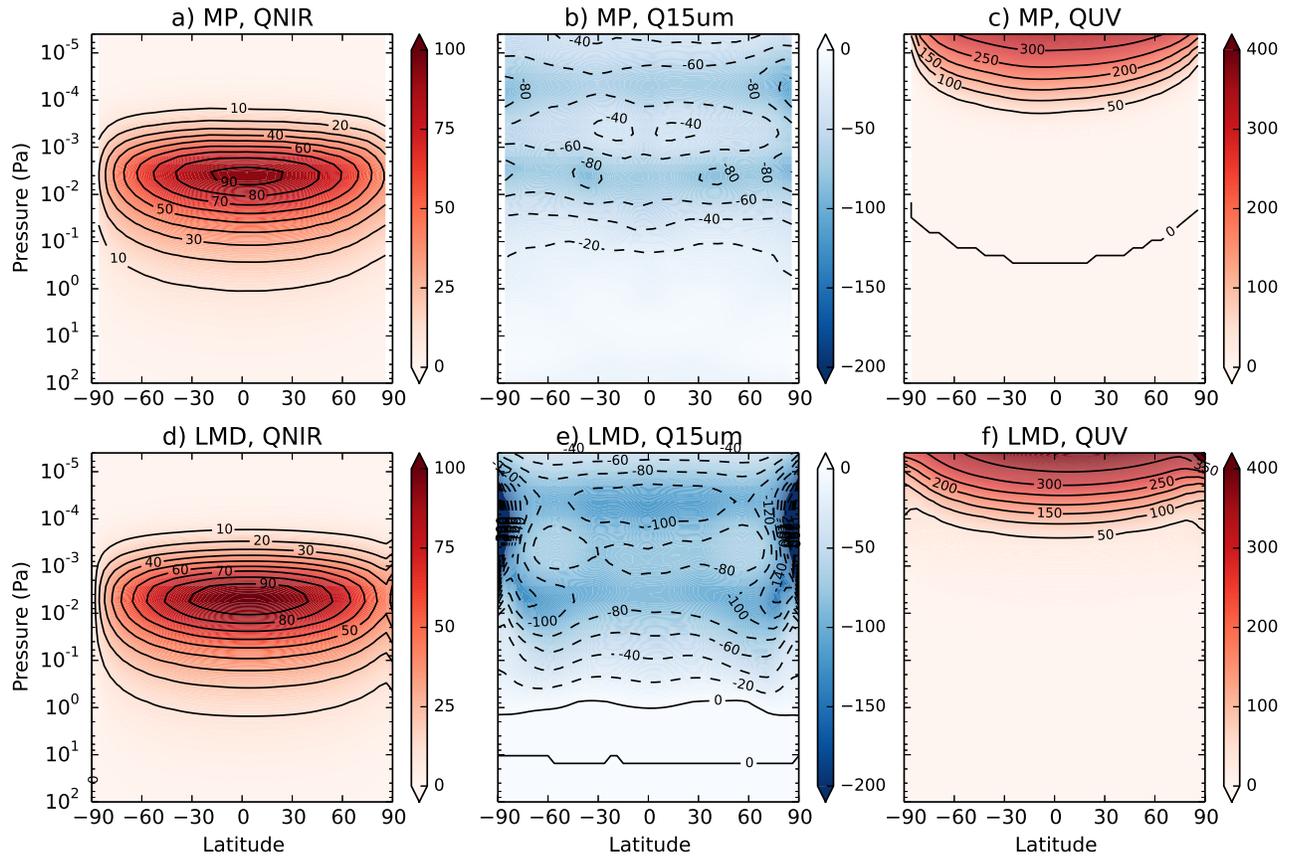}
\caption{Radiative heating/cooling rates (in K~sol$^{-1}$) from the runs
with the MCD-1D oxygen profile without including gravity wave effects:
for MPI--GCM (upper row), and LMD--MGCM (lower row). The left column is due
to the near-IR CO$_2$ heating, the center column is due the 15 $\mu$m band
CO$_2$ cooling, and the right column is for the UV--EUV heating.}
\label{fig:1Dqterms}
\end{figure}

Further insight can be gained from analyzing the radiative heating/cooling
rates, which are plotted for the MCD-1D runs in Figure~\ref{fig:1Dqterms}.
Both GCMs employ the same parameterization for near-IR CO$_2$ heating, and,
therefore, the calculated mean rates are within the margins of errors due to
numerics (Figure~\ref{fig:1Dqterms}a,d). The maximum of heating ($\approx$90
K~sol$^{-1}$) occurs just below the mesopause, and steeply decreases above.
On the contrary, the UV-EUV heating rates continuously increase above the
mesopause. This heating rate is also similar in both simulations (with a
somewhat stronger heating in the LMD--MGCM) despite the completely independent
parameterizations used in the two models (Figure~\ref{fig:1Dqterms}c,f).
Cooling rates due to the 15 $\mu$m CO$_2$ band differ the most. Unlike the EUV
and near-IR heating rates, they strongly depend on temperature. In both models,
CO$_2$ cooling rates increase with height in the mesosphere, have local minima
at the mesopause (as a response to temperature minima), reach maxima above the
mesopause, and decay with height in the lower thermosphere
(Figure~\ref{fig:1Dqterms}b,e). Cooling is stronger in high latitudes in both
GCMs, which is  a consequence of higher polar temperatures. However, CO$_2$
cooling rates in the LMD model are overall larger than in the MPI--GCM,
especially near the poles: up to --400 K~sol$^{-1}$ poles in the LMD run versus
$-\sim$110 K~sol$^{-1}$ in the MPI simulation. In the regions where the model
temperatures are closer, the corresponding cooling rates are closer as well,
for instance, around --80 K~sol$^{-1}$ at $p=0.01$ Pa in low latitudes.
Qualitatively, both models demonstrate a similar response to variations of
atomic oxygen with larger cooling rates in the LMD model.

\subsection{Impact of Gravity Waves}

\begin{figure}
\centering
\includegraphics[width=1.0\columnwidth,clip]{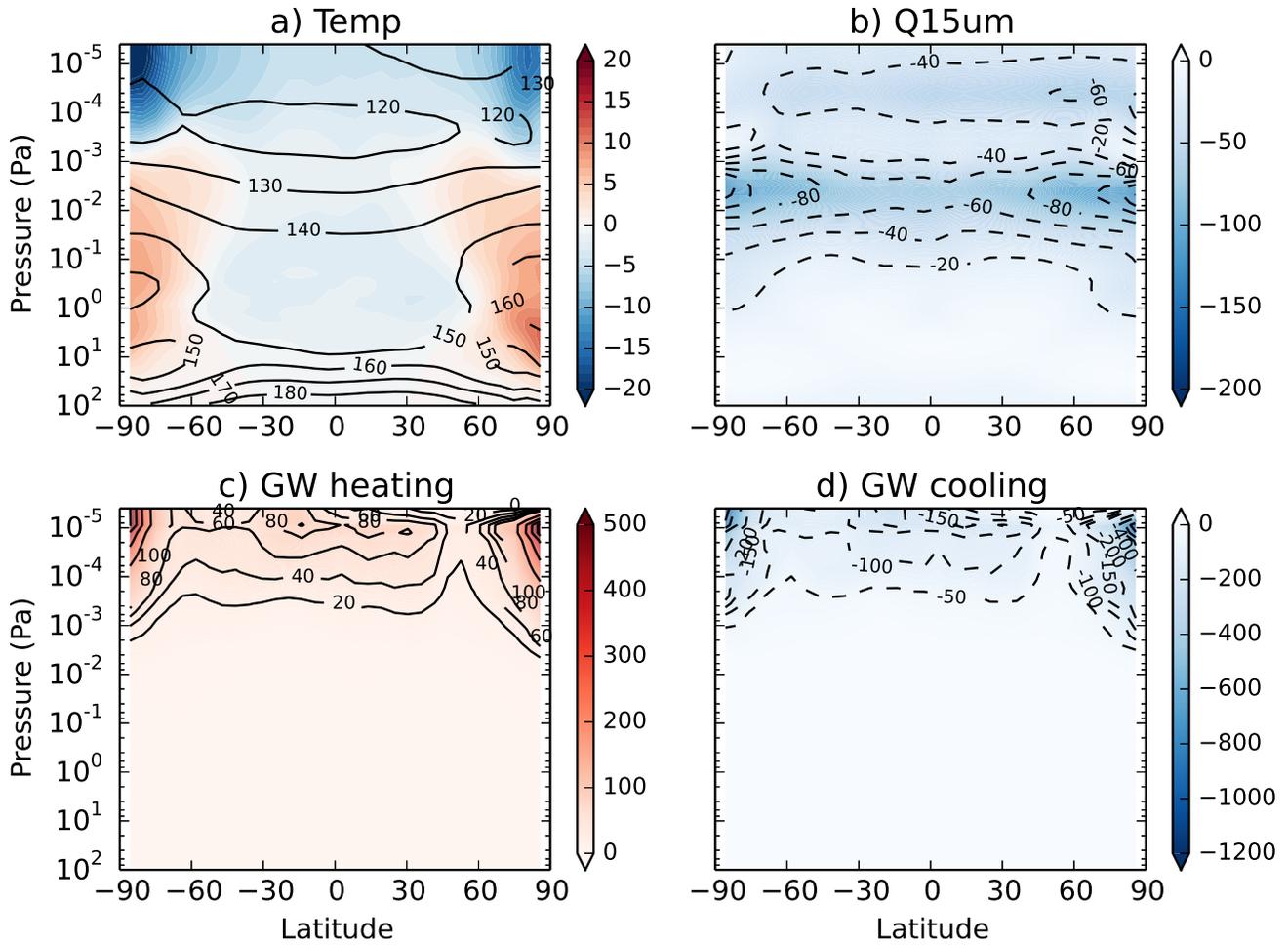}
\caption{Results of simulations with MPI--GCM for the MCD-1D oxygen scenario.
(a) Temperature (contours) and temperature difference with the MCD-1D run
without the parameterized gravity wave effects (shaded).
(b) 15 $\mu$m CO$_2$ cooling rates (contours and shaded).
(c) GW-induced heating, and (d) cooling rates.}
\label{fig:1Dgw}
\end{figure}

After exploring the model responses to enhanced CO$_2$ cooling due to the
prescribed atomic oxygen, we turn to thermal effects of gravity waves. For
that, we included the GW scheme in the MPI--GCM, and repeated the simulations.
Results for the MCD-1D scenario are plotted in Figure~\ref{fig:1Dgw}.
Two distinctive effects of GWs are seen in the temperature field
(Figure~\ref{fig:1Dgw}a). They are (a) the colder lower thermosphere,
especially in high latitudes, where the simulated temperature dropped by up
to 20 K, and (b) warmer (by up to 9 K) polar regions in the middle atmosphere.
In the run with the ``low" atomic oxygen scenario (not shown here), the
GW-induced changes in the thermosphere were even stronger (more than 30 K
colder poles in the thermosphere), but qualitatively similar. Such effects of
gravity waves have previously been reported for solstice conditions
\citep{MY12}, and then throughout the Martian year \citep[Figure~1]{mykh13}.

As a response to the altered simulated temperature, CO$_2$ cooling rates in
Figure~\ref{fig:1Dgw}b have also changed with respect to those in
Figure~\ref{fig:1Dqterms}b for the run without GWs. In particular, peak values
increased from --80 to --120 K~sol$^{-1}$ over warmer polar regions in the
mesosphere, and decreased from --100 to --60 K~sol$^{-1}$ in the colder lower
thermosphere. GW-induced heating and cooling rates in Figure~\ref{fig:1Dgw}c,d
can now be compared against those due to radiative transfer in CO$_2$
molecules. Firstly, the former takes place, generally, higher than the latter.
Secondly, the magnitudes of GW heating and cooling ($E^j$ and $Q^j$ in
Equation \ref{eq:gw-heat}, correspondingly), are comparable or larger than 
those by CO$_2$. Thirdly, GW heating and cooling increase with height in the 
lower thermosphere, whereas the radiative rates reach their maxima there, and 
then decay with altitude. GW cooling exceeds heating everywhere in the lower 
thermosphere. Thus, the net effect of GWs is $\sim$--100 K~sol$^{-1}$ at 
$p=10^{-5}$ Pa in low and middle latitudes, and is more than $\sim$--600
K~sol$^{-1}$ near the poles at these heights. The enhancement of
cooling/heating rates at high-latitudes is related to the stronger GW activity 
in polar regions, which results from more favorable propagation conditions for 
GW harmonics there. Such increased activity has been found in the thermosphere 
of Earth at solstices and equinoxes \citep{Yigit_etal09,Yigit_etal12, 
Yigit_etal14}, and of Mars throughout most of the year \citep{mykh13}. Note 
that higher temperatures in the polar regions in the mesosphere are not 
related to direct heating by GWs, as is seen in Figures~\ref{fig:1Dgw}c,d. 
They are the result of the adiabatic heating produced by the enhanced 
downward branch of the meridional circulation driven dynamically by GWs.

\section{Results for Two-Dimensional Atomic Oxygen Distribution}
\label{sec:2D}

\begin{figure}
\vspace{-1cm}
\centering
\includegraphics[width=0.45\columnwidth,clip]{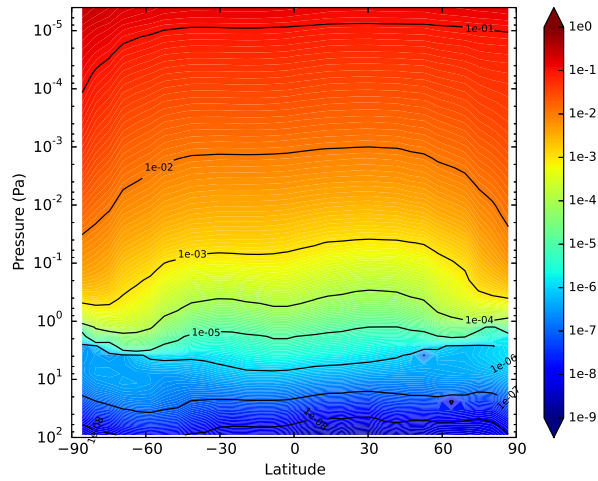}
\caption{Latitude-height distribution of atomic oxygen volume mixing ratios
for the MCD-2D scenario.}
\label{fig:2DOx}
\end{figure}

Distribution of atomic oxygen in the lower thermosphere is controled by
photochemistry and transport. During equinoxes, the two-cell circulation with
downward branches over the poles creates an excess of [O] in high latitudes.
In this section, we quantify the effects of latitudinal variations of atomic
oxygen. For that, we again use the MCD output for $L_s=0$--30$^\circ$ averaged
over longitudes and local times. This scenario (referred to as MCD-2D) with
the atomic oxygen varying with latitude and height is consistent with the
MCD-1D, as the latter is obtained by averaging the former. As seen from the
corresponding altitude-height distribution in Figure~\ref{fig:2DOx}, there is
up to 5 times more atomic oxygen in polar regions than in the low- and middle
latitudes, with somewhat more over the South Pole. Because of that, there is
less [O] in low- and middle-latitudes in the MCD-2D scenario than in the MCD-1D.

\begin{figure}
\vspace{-1cm}
\centering
\includegraphics[width=0.33\columnwidth,angle=-90,clip]{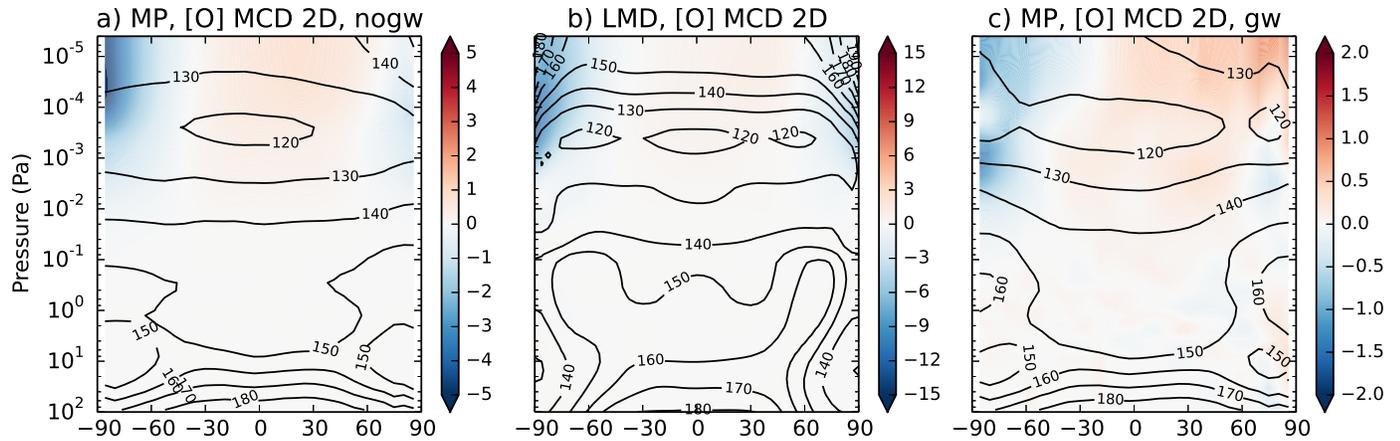}
\caption{Temperature simulated for the MCD-2D scenario (contour lines):
(a) by the MPI--MGCM without parameterized GW effects; (b) by the LMD-MGCM
without GW effects; (c) by the MPI--GCM with GW parameterization included.
Color shades denote temperature differences with the corresponding runs
for the MCD-1D scenario.}
\label{fig:2D}
\end{figure}

The simulated MCD-2D temperatures are plotted in Figure~\ref{fig:2D} with
contours, while the color shades denote the differences with respect to the
MCD-1D runs. In full accordance with the atomic oxygen distribution, most of
the changes occurred in the polar regions of the thermosphere, with more in
the Southern Hemisphere. This difference reaches up to --15 K in the LMD--MGCM
simulation (Figure~\ref{fig:2D}b), while it is more moderate (up to --5 K,
Figure~\ref{fig:2D}a) in the MPI--GCM run without GWs. In the low and middle
latitudes, the simulated temperatures are subtly higher ($\sim$1 K) as there
is less [O] there. Oxygen-induced temperature change is smaller, when runs
with GWs are compared (Figure~\ref{fig:2D}c). In the Southern Hemisphere it is
only $\sim$--2 K, and even somewhat positive ($\sim$1 K) in the Northern
high-latitudes. However the absolute values of temperature at and above the
mesopause remain lower with GWs. Clearly, their thermal effects dominate in
these regions.

Overall, our simulations show that the latitudinal variations of atomic oxygen
do affect the simulated temperature in the lower thermosphere, however, the
changes they induce are smaller than those introduced by vertical variations. 
On the one hand, this is because the atomic oxygen abundances differ more
between the MCD-1D and \citet{Nair94} scenarios than between the MCD-1D and
MCD-2D. On the other hand, latitudinal gradients of [O] and of the related
adiabatic heating/cooling rates alter the meridional circulation, and the
associated adiabatic heating.

\section{Net Diabatic Heating/Cooling}
\label{sec:net}

\begin{figure}
\vspace{-1cm}
\centering
\includegraphics[width=0.80\columnwidth,angle=-90,clip]{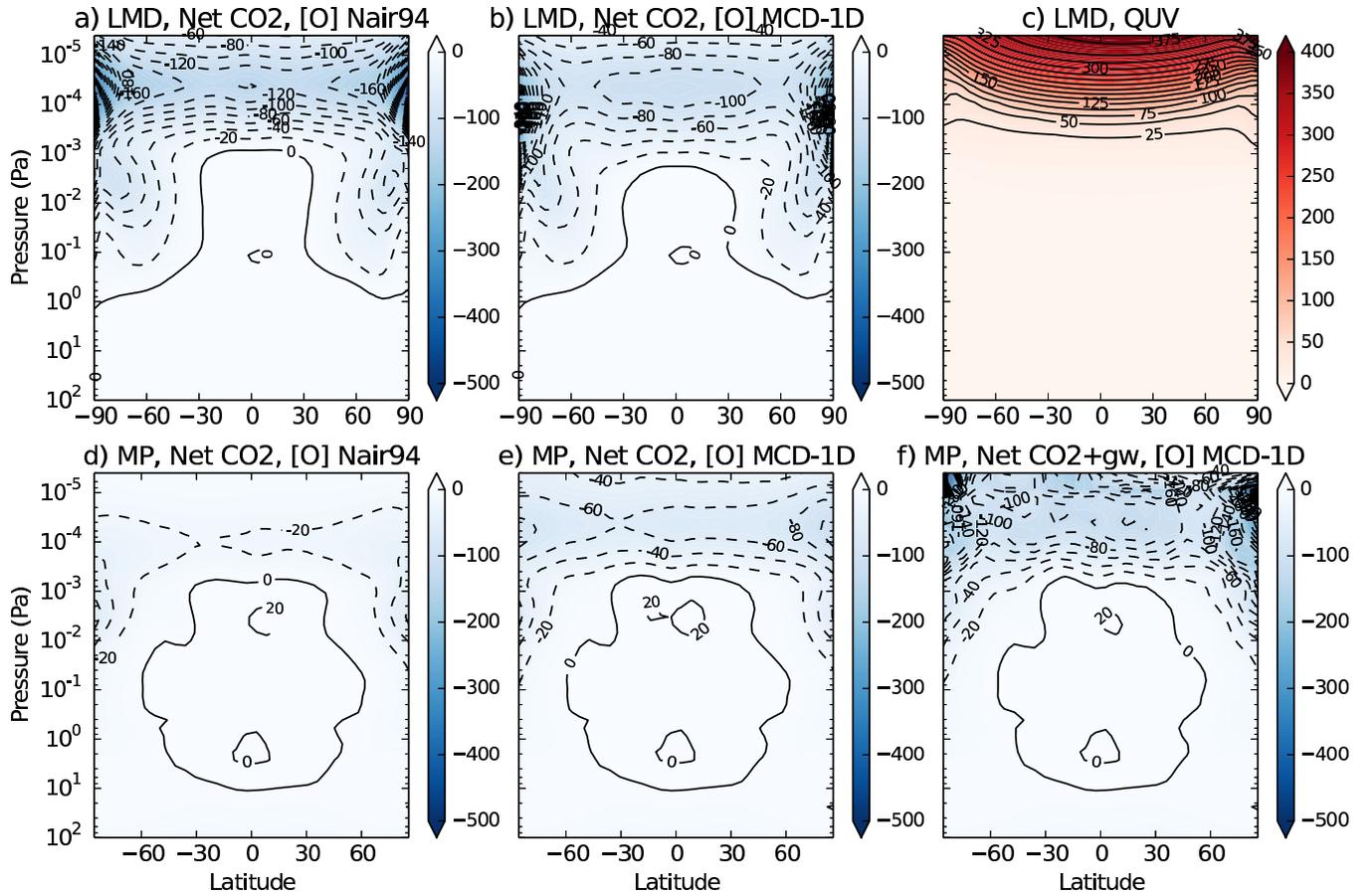}
\caption{Net heating/cooling rates for the scenario with one-dimensional
atomic oxygen profiles. Upper row is from the LMD-MGCM simulations, and
the lower row is from the MPI--GCM.
a) Net CO$_2$ radiative rates (15 $\mu$m cooling
and NIR heating) for the \citet{Nair94} scenario;
b) same as (a), but for the MCD-1D scenario;
c) UV-EUV heating;
d) same as (a), but for the MPI--GCM;
e) same as in (b), but for the MPI--GCM;
f) total of CO$_2$ radiative and gravity wave heating and cooling rates.}
\label{fig:net}
\end{figure}

In order to explore the interplay between the radiative and gravity wave 
forcings, and the global dynamics further, we plotted the net diabatic rates 
(with the exception of UV-EUV heating, which remains constant in all the 
simulations) in Figure~\ref{fig:net}. The UV-EUV heating is also shown 
separately in the panel c. For simplicity, we consider only one-dimensional 
(vertically varying) [O] scenarios of \citet{Nair94} and MCD-1D. All the 
simulations (Figure~\ref{fig:net}a,b,d,e,f) show the net diabatic heating in 
low latitudes up to approximately mesopause level, which is created by the 
CO$_2$ NIR heating, and cooling in middle, and especially, high latitudes. 
In the mean sense (when the transience is neglected), this radiative forcing 
is compensated by the adiabatic cooling and heating associated with the rising
and sinking branches of the meridional circulation, respectively. Thus, it is
seen that the mean meridional circulation in the mesosphere consists of two
cells with the upwelling in low latitudes, downwelling in high latitudes, and
the poleward transport in both hemispheres. Hence, the resulting temperature 
is the function of a delicate balance between the diabatic and adiabatic 
forcing.

The 15-$\mu$m CO$_2$ band is the major contributor to the diabatic cooling. 
It depends on temperature, and, indirectly, on the meridional circulation as
well. In the LMD model, these cooling rates are, in general, larger than those
in the MPI--MGCM, especially in high latitudes. For the MCD-1D (``large" oxygen)
scenario, the MPI simulation shows an overall increase of the CO$_2$ cooling
rates by up to a factor of 5 near the top (Figure~\ref{fig:net}d,e).  
The response of the LMD--MGCM run is less straightforward: the cooling rates 
became even somewhat weaker in most parts in the domain despite the greater 
amount of atomic oxygen (Figure~\ref{fig:net}a,b). This example demonstrates
that the model response to varying [O] amount is highly non-linear, and driven
not merely by the local diabatic forcing, but by the dynamics as well.

Gravity wave effects add to the CO$_2$ cooling (Figure~\ref{fig:net}f), and 
the total diabatic cooling rates in the MPI--GCM run turn out to be very 
similar to those from the LMD--MGCM simulation without GW effects. 
Thus, gravity waves contribute to the complexity of interactions between the 
diabatic forcing and dynamics in the atmosphere. Note also that GW cooling 
acts against the UV-EUV heating in the upper portion of the model domain, and, 
thus, adds to the main cooling mechanism in the thermosphere -- molecular heat 
conduction.

\section{CO$_2$ Density}
\label{sec:dens}

Having explored the influence of CO$_2$ and GW cooling on temperature, we
now turn to the neutral density, to which CO$_2$ is the major contributor in 
the lower thermosphere. Densities at these heights impact aerobraking
operations, and, therefore, their quantification is of great interest.

\begin{figure}
\vspace{-1cm}
\centering
\includegraphics[width=0.72\columnwidth,angle=-90,clip]{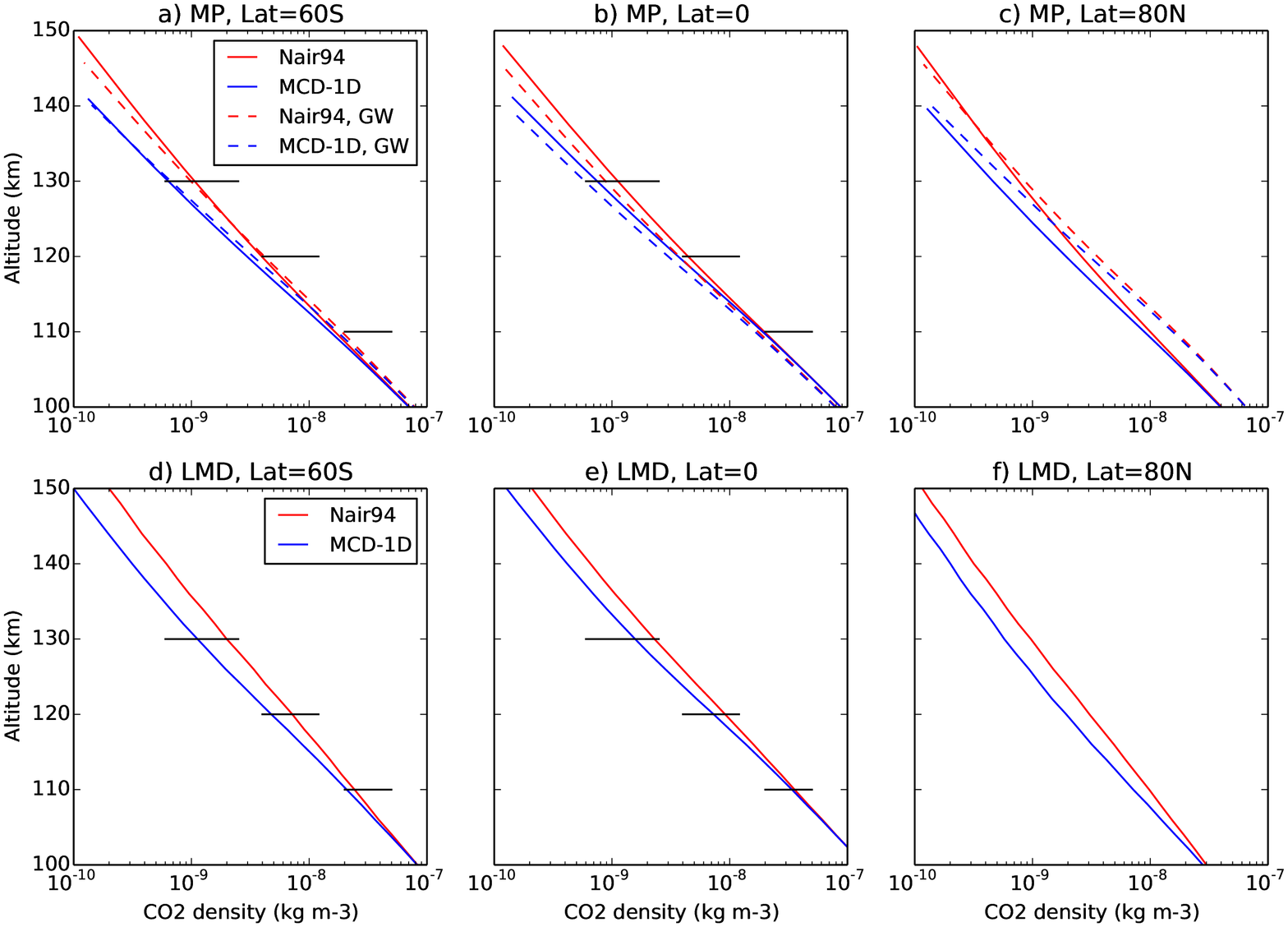}
\caption{Mean neutral (CO$_2$) density simulated with the MPI--MGCM (upper row) 
and the LMD--MGCM (lower row) at 60$^\circ$S (left column), over the equator 
(central column), and at 80$^\circ$N (right column).
Red lines correspond to the ``Nair94" atomic oxygen scenario, blue lines 
are for the MCD-1D scenario. Solid and dashed lines denote simulations 
without and with GW parameterization included, respectively. Black horizontal
lines represent the range of density variability from the SPICAM observations.}
\label{fig:dens}
\end{figure}

The effects of atmospheric temperature on density are two-fold. On the one 
hand, density is inversely proportional to temperature in accordance with the 
equation of state. Hence, colder thermospheric air at a given model pressure 
level implies its higher density. On the other hand, lower temperature 
means the smaller density scale height, and, therefore, the lower geometrical 
altitude of the pressure level. Thus, the net density response to the colder 
simulated thermosphere is not straightforward, but represents an interplay of 
these two effects. The height of the pressure levels (``geopotential height") 
is the prognostic field in the MGCMs, and we used it to calculate the modeled
densities as functions of geometrical altitudes. Figure~\ref{fig:dens} shows
the vertical profiles of mean density from the simulations with the two MGCMs
at three characteristic latitudes: a) at middle-to-high latitudes 
(60$^\circ$S), b) over the equator, and c) in the polar region (80$^\circ$N). 
Solid lines denote the simulations without GWs, and the dashed lines (in the 
MPI--MGCM simulations) are for runs with the GW scheme included.
Between 110 and 130 km, where aerobraking operations take place, densities
in the larger-oxygen MCD-1D scenario (blue lines) are up to 20$\%$ smaller
than in the corresponding ``Nair94" runs (red lines) in both models. Higher, 
the difference increases to 100\% and grows with height in the MPI--GCM, while 
remains almost constant with height above about 130 km in the LMD--MGCM. This 
is consistent with the modifications of the thermal structure induced by 
oxygen changes in both models, mostly focused on the mesopause for the 
LMD--MGCM and being more important in the lower thermosphere for the MPI--GCM 
(see Figure~\ref{fig:1D-old}). Gravity wave-induced cooling provides an
additional contribution to the density decrease in the MPI--GCM simulations.

At low- and middle latitudes, most of temperature changes occur at the
mesosphere and higher, and the densities do not differ between simulations
below $\approx$100 km (Figure~\ref{fig:dens}a,b,d,e). The complex dependence 
of density on the atmospheric temperature is particularly seen at high 
latitudes (Figure~\ref{fig:dens}c,f), where GWs enhance the middle atmosphere 
polar warming, thus altering densities at much lower altitudes. At 80$^\circ$
around 100 km, the density simulated with accounting for GWs is $\sim$20\% 
larger in both oxygen scenarios. Higher, the vertical behavior is consistent 
with that shown in panels a and b.

Finally, we compare the simulated densities with the SPICAM measurements.
The latter are presented in Figure~4 of the paper by \citet{forg09} as 
functions of the solar longitude $L_s$ at different heights for latitudes 
between 50$^\circ$N and 50$^\circ$S. We plotted the range of their variations 
near $L_s\approx 0^\circ$ with horizontal bars in Figure~\ref{fig:dens}a,b,d,e. 
At 130 km, the SPICAM measurements are around $10^{-9}$ kg~m$^{-3}$. At 120 km, 
they are scattered between $4\times 10^{-9}$ and $10^{-8}$ kg~m$^{-3}$. At 110 
km, the observed densities are between 2 and $5\times 10^{-8}$ kg~m$^{-3}$. 
It is seen that the MPI densities are on the lower end of measurements, while 
those from the LMD runs are somewhat larger and reproduce nicely the SPICAM 
measured densities, in particular for the simulations with the MCD-1D oxygen 
profile. Given that the simulated densities are the averaged quantities, and 
the SPICAM observations have been taken at certain local times being affected 
by thermal tides, which are particularly strong in the thermosphere, the 
agreement can be considered as quite good in both cases.

\section{Conclusions}
\label{sec:conc}

We explored cooling effects caused by CO$_2$ radiation and gravity waves
(GWs) in the Martian mesosphere and lower thermosphere using two Martian
general circulation models (GCMs), the Laboratoire de M\'et\'eorologie 
Dynamique (LMD) and Max Planck Institute (MPI) GCMs, for equinox conditions 
around $L_s=0^{\circ}$ and low dust and solar activity. The main inferences 
of our simulations are listed below.

\begin{enumerate}
\item
Within the present day uncertainties with distributions of atomic oxygen
and gravity wave sources, both CO$_2$ and GW cooling can compensate for 10 to 
40 K warmer model temperature bias in the lower thermosphere compared to
observations, and produce a colder simulated mesopause.

\item
CO$_2$ radiation under larger atomic oxygen abundances, and GWs impose 
stronger cooling on the Martian lower thermosphere, and result in colder model 
temperatures at high latitudes, especially by GWs under a low oxygen scenario.

\item
Gravity wave cooling takes place, generally, higher than that due to CO$_2$
radiation, and GW cooling rates increase with height into the upper
thermosphere, as opposed to the CO$_2$ rates that peak in the lower
thermosphere.

\item
Because most of GW activity occurs at middle and high latitudes of the
thermosphere, the cooling due to GWs dominates there. The oxygen-induced 
amplification of the CO$_2$ cooling is dominant at low latitudes.

\item
In addition to cooling in the thermosphere, GWs enhance the polar warmings
in the mesosphere, unlike the CO$_2$ radiation. This temperature increase is
produced dynamically (through the adiabatic heating associated with the
downward branches of the meridional circulation), rather than thermally.

\item
The simulated atmospheric densities at the typical altitudes for aerobraking 
operations decrease in around 20\% at all latitudes with the larger atomic 
oxygen abundances. Gravity waves activity further modifies the densities in 
the polar regions. The simulated densities are in overall agreement with the
SPICAM measurements, in particular when the larger atomic oxygen profile is 
used in the simulations.
\end{enumerate}

During high solar activity, the amount of atomic oxygen in the lower
thermosphere increases. Likewise, gravity waves penetrate higher into the
thermosphere during the active Sun periods, albeit their effects are generally
weaker than during low solar activity \citep{YM10}. Therefore, our conclusions 
may change for such conditions.
What are the possible pathways for further understanding the role of CO$_2$
and GW cooling, and eliminating the warm model bias in the lower thermosphere?
Firstly, this can be done by constraining the parameterizations. For that,
measurements of atomic oxygen volume mixing ratios in the thermosphere, and of
GW sources in the lower atmosphere are required. Both tasks are not trivial,
and cannot be performed immediately. Secondly, more detailed temperature
measurements could provide the required information.
In particular, latitudinal gradients throughout the thermosphere and
temperature above $\sim$110 km could help to clarify the role of GWs.
Such data will soon become available from the Imaging UltraViolet Spectrograph
(IUVS) and Neutral Gas and Ion Mass Spectrometer (NGIMS) onboard the
operating MAVEN (Mars Atmosphere and Volatile EvolutioN) orbiter.


\begin{acknowledgments}
Mars Climate Database is available at \\
http://www-mars.lmd.jussieu.fr/mars/access.html. Data supporting the 
figures are available upon request from ASM (medvedev@mps.mpg.de) and FGG
(ggalindo@iaa.es).

The work was partially supported by German Science Foundation (DFG) grant
ME2752/3-1. FGG was funded by a CSIC JAE-Doc contract cofinanced by the 
European Social Fund. FGG thanks the Spanish MICINN for funding support 
through the CONSOLIDER program ASTROMOL CSD2009-00038, and through project 
AYA2011-23552/ESP. EY was partially supported by NASA grant NNX13AO36G.
\end{acknowledgments}


%
%
\end{article}

\newpage

\begin{table}
\caption{Approximate Height Versus Pressure}
\setlength{\tabcolsep}{10pt}
\centering
   \begin{tabular}{|c|c|c|c|c|c|c|c|c|}
   \hline
   \noalign{\smallskip}
Pressure, Pa &  100  &  10   &    1    &   0.1  &  0.01  &  0.001  &  0.0001  &
 $10^{-5}$  \\
   \noalign{\smallskip} \hline \noalign{\smallskip}
Height, km   &  18.3 & 38.0  &  57.5   &   75.2 &  98.6  &  108.0  &   124.1  &
 141.2  \\
   \noalign{\smallskip}
   \hline
   \end{tabular}
\label{tab:1}
\end{table}

\begin{thebibliography}{}

\providecommand{\natexlab}[1]{#1}
\expandafter\ifx\csname urlstyle\endcsname\relax
  \providecommand{\doi}[1]{doi:\discretionary{}{}{}#1}\else
  \providecommand{\doi}{doi:\discretionary{}{}{}\begingroup
  \urlstyle{rm}\Url}\fi

\bibitem[{\textit{Angelats i Coll et al.}(2005)}]{ange05}
Angelats i Coll, M., F. Forget, M. A. L\'{o}pez-Valverde, and
F. Gonz\'{a}lez-Galindo (2005), The first Mars thermospheric general
circulation model: The Martian atmosphere from the ground to 240 km,
\textit{Geophys. Res. Lett.}, \textit{32}, L04201, doi:10.1029/2004GL021368.

\bibitem[{\textit{Bell et al.}(2007)}]{bell07}
Bell, J. M., S. W. Bougher, and J. R Murphy (2007), Vertical dust mixing
and the interannual variations in the Mars thermosphere,
\textit{J. Geophys. Res.}, \textit{112}, E12002, doi:10.1029/2006JE002856.

\bibitem[{\textit{Bertaux et al.}(2006)}]{bert06}
Bertaux, J.-L., et al.
(2006),
SPICAM on Mars Express: Observing modes and overview of UV spectrometer data
and scientific results, \textit{J. Geophys. Res.}, \textit{111}, E10S90,
doi:10.1029/2006JE002690.

\bibitem[{\textit{Bougher et al.}(2009)}]{bough09}
Bougher, S. W., T.~M. McDunn, K.~A. Zoldak, and J.~M. Forbes (2009), Solar
cycle variability of Mars dayside exospheric temperatures: MTGCM evaluation
of underlying thermal balances, \textit{Geophys. Res. Lett.}, \textit{36},
doi:10.1029/2008GL036376.

\bibitem[{\textit{Bougher et al.}(2006)}]{bough06}
Bougher, S.~W., J.~M. Bell, J.~R. Murphy, M.~A. Lopez-Valverde,
and P.~G. Withers (2006), Polar warming in the Mars thermosphere:
Seasonal variations owing to changing insolation and dust distributions,
\textit{Geophys. Res. Lett.}, \textit{33}, L02203, doi:10.1029/2005GL024059.

\bibitem[{\textit{Creasey et al.}(2006)}]{creasey06}
Creasey, J.~E., J.~M. Forbes, and D.~P. Hinson (2006), Global and seasonal
distribution of gravity wave activity in Mars' lower atmosphere derived
from MGS radio occultation data, \textit{Geophys. Res. Lett.},
\textit{33}, L01803, doi:10.1029/2005GL024037.

\bibitem[{\textit{Conrath}(1975)}]{conr75}
Conrath, B. J., (1975), Thermal structure of the Martian atmosphere during the
dissipation of the dust storm of 1971, \textit{Icarus}, \textit{24}, 36--46.

\bibitem[{\textit{Feofilov et al.}(2006)}]{feofilov2006}
Feofilov, A.~G., A.~A.~Kutepov, A.~S.~Medvedev, and P.~Hartogh (2012),
New technique for calculating the non-LTE infrared radiative cooling/heating
rates in the Martian GCM, Second workshop on Mars atmosphere modelling and
observations, 27.02-03.03, 2006 Granada, Spain, Edited by F.~Forget,
M.~A.~Lopez-Valverde, M.~C.~Desjean, J.~P.~Huot, F.~Lefevre, S.~Lebonnois,
S.~R.~Lewis, E.~Millour, P.~L.~Read and R.~J.~Wilson.
\textit{Publisher: LMD, IAA, AOPP, CNES, ESA}, 614--617.

\bibitem[{\textit{Feofilov et al.}(2012)}]{feofilov2012}
Feofilov, A.~G., A.~A.~Kutepov, C.-Y.~She, A.~K.~Smith, W.~D.~Pesnell, and
R.~A.~Goldberg (2012), CO$_2$($\nu_2$)-O quenching rate coefficient derived
from coincidental SABER/TIMED and Fort Collins lidar observations of the
mesosphere and lower thermosphere, \textit{Atmos. Chem. Phys.}, \textit{12},
9013--9023.

\bibitem[{\textit{Feofilov and Kutepov}(2012)}]{feofilovkutepov2012}
Feofilov, A.~G. and A.~A.~Kutepov (2012), Infrared radiation in the mesosphere
and lower thermosphere: energetic effects and remote sensing,
\textit{Surv. Geophys.}, doi:10.1007/s10712-012-9204-0.

\bibitem[{\textit{Forget et al.}(2009)}]{forg09}
Forget, F., F.~Montmessin, J.-L. Bertaux, F. Gonz\'{a}lez-Galindo,
S.~Lebonnois, E.~Qu\'{e}merais, A.~Reberac, E.~Dimarellis, and
M.~A. L\'{o}pez-Valverde (2009), Density and temperatures of the upper
Martian atmosphere measured by stellar occultations with Mars Express SPICAM.
\textit{J. Geophys. Res.}, \textit{114}, E01004, doi:10.1029/2008JE003086.

\bibitem[{\textit{Forget et al.}(1999)}]{forg99}
Forget, F., F. Hourdin, R. Fournier, C. Hourdin, O. Talagrand,
M. Collins, S.~R.~Lewis, P.~L. Read, J.-P. Huot (1999), Improved
general circulation models of the Martian atmosphere from the surface
to above 80 km,  \textit{J. Geophys. Res.}, \textit{104}, 24155--24175.

\bibitem[{\textit{Fox et al.}(1996)}]{fox96}
Fox, J. L., P. Zhou, and S. W. Bougher (1996),
The Martian thermosphere/ionosphere at high and low solar activities,
\textit{Adv. Space Res.}, \textit{17}, 203--218.

\bibitem[{\textit{Gonz\'{a}lez-Galindo et al.}(2013)}]{gonz13}
Gonz\'{a}lez-Galindo, F., J.-Y. Chaufray, M.~A. L\'{o}pez-Valverde, G. Gilli,
F. Forget, F. Leblanc, R. Modolo, S. Hess, and M. Yagi (2013),
Three-dimensional Martian ionosphere model: I. The photochemical ionosphere
below 180 km, \textit{J. Geophys. Res.}, \textit{118}, 2105--2123,
doi:10.1002/jgre.20150.

\bibitem[{\textit{Gonz\'{a}lez-Galindo et al.}(2010)}]{gonz10}
Gonz\'{a}lez-Galindo, F., S. W. Bougher, M.~A. L\'{o}pez-Valverde, F. Forget,
and J. Murphy (2010), Thermal and wind structure of the Martian thermosphere
as given by two general circulation models, \textit{Planet. Space Sci.},
\textit{58}, 1832--1840.

\bibitem[{\textit{Gonz\'{a}lez-Galindo et al.}(2009a)}]{gonz09}
Gonz\'{a}lez-Galindo, F., F. Forget, M.~A. L\'{o}pez-Valverde,
M.~Angelats i Colli, and E.~Millour (2009a), A ground-to-exosphere
Martian general circulation model: 1. Seasonal, diurnal, and solar
cycle variation of thermospheric temperatures, \textit{J. Geophys. Res.},
\textit{114}, E04001, doi:10.1029/2008JE003246.

\bibitem[{\textit{Gonz\'{a}lez-Galindo et al.}(2009b)}]{gonz09b}
Gonz\'{a}lez-Galindo, F., F. Forget, M.~A. L\'{o}pez-Valverde, and
M.~Angelats i Colli (2009b), A ground-to-exosphere
Martian general circulation model: 2. Atmosphere during solstice
conditions - Thermospheric polar warming, \textit{J. Geophys. Res.},
\textit{114}, E08004, doi:10.1029/2008JE003277.

\bibitem[{\textit{Gonz\'{a}lez-Galindo et al.}(2005)}]{gonz05}
Gonz\'{a}lez-Galindo, F., M. A. L\'{o}pez-Valverde, M. Angelats i Coll,
and F. Forget (2005), Extension of a Martian general circulation model to
thermospheric altitudes: UV heating and photochemical models, J. Geophys.
Res., 110, E09008, doi:10.1029/2004JE002312.

\bibitem[{\textit{Gusev and Kutepov}(2003)}]{gusev03}
Gusev, O.~A., and A.~A. Kutepov (2003), Non-LTE gas in planetary atmospheres,
in \textit{Stellar Atmosphere Modeling}, edited by I. Hubeny, D. Mihalas,
and K. Werner, \textit{ASP Conference Series}, \textit{288}, 318--330.

\bibitem[{\textit{Hanson et al.}(1977)}]{hanson77}
Hanson, W. B., S. Santanini, and D. R. Zuccaro (1977), The Martian ionosphere
as observed by Viking retarding potential analyzers,
\textit{J. Geophys. Res.}, \textit{82}, 4351--4363.

\bibitem[{\textit{Hartogh et al.}(2007)}]{hart07}
Hartogh, P., A.~S. Medvedev, and C. Jarchow (2007), Middle atmosphere polar
warmings on Mars: simulations and study on the validation with sub-millimeter
observations, \textit{Planet. Space Sci.}, \textit{55}, 1103--1112.

\bibitem[{\textit{Hartogh et al.}(2005)}]{hart05}
Hartogh, P., A.~S. Medvedev, T.~Kuroda, R. Saito, G.~Villanueva,
A.~G.~Feofilov, A.~A.~Kutepov, and U.~Berger (2005), Description and
climatology of a new general circulation model of the Martian
atmosphere, \textit{J. Geophys. Res.}, \textit{110}, E11008,
doi:10.1029/2005JE002498.

\bibitem[{\textit{Huestis et al.}(2008)}]{huestis08}
Huestis, D. L., S.~W. Bougher, J.~L. Fox, M.~Galand, R.~E. Johnson,
J.~I. Moses, and J.~C. Pickering (2008), Cross sections and reaction rates
for comparative planetary aeronomy, \textit{Space Sci. Rev.},
\textit{139}, 63--105.

\bibitem[{\textit{Keating et al.}(1998)}]{Keating98}
Keating, G.~M., et al. (1998), The structure of the upper atmosphere of
Mars: In-situ accelerometer measurements from Mars Global Surveyor,
\textit{Science}, \textit{279}, 1672--1676.

\bibitem[{\textit{Kutepov et al.}(1998)}]{kutepov98}
Kutepov, A.~A., O.~A. Gusev, and V.~P. Ogibalov (1998), Solution of the
non-LTE problem for molecular gas in planetary atmospheres:
Superiority of accelerated lambda iteration,
\textit{J. Quant. Spectrosc. Radiat. Transf.}, \textit{60}, 199--220.

\bibitem[{\textit{Lef\`evre et al.}(2004)}]{lefe04}
Lef\`{e}vre, S., S. Lebonnois, F. Montmessin, and F. Forget (2004), Three
dimensional modeling of ozone on Mars, \textit{J. Geophys. Res.},
\textit{109}, E07004, doi:10.1029/2004JE002268.

\bibitem[{\textit{L\'{o}pez-Valverde and L\'{o}pez-Puertas}(2001)}]{lval01}
L\'{o}pez-Valverde, M. A., and M. L\'opez-Puertas (2001), Atmospheric
non-LTE effects and their parameterization for Mars, ESA, Technical Report.

\bibitem[{\textit{McDunn et al.}(2010)}]{mcdunn10}
McDunn, T.~L., S.~W. Bougher, J. Murphy, M.~D. Smith, F. Forget,
J.-L. Bertaux, and F. Montmessin (2010), Simulating the density and thermal
structure of the middle atmosphere (< 80--130 km) of Mars using the
MGCM-MTGCM: a comparison with MEX/SPICAM observations, \textit{Icarus},
\textit{206}, 5--17.

\bibitem[\textit{Medvedev and Yi\u{g}it}(2012)]{MY12}
Medvedev,~A.~S., and E.~Yi\u{g}it (2012), Thermal effects of internal gravity
waves in the Martian thermosphere, \textit{Geophys. Res. Lett.},
\textit{39}, L05201, doi:10.1029/2012GL050852.

\bibitem[{\textit{Medvedev et al.}(2013)}]{mykh13}
Medvedev,~A.~S., E.~Yi\u{g}it, T. Kuroda, and P.~Hartogh (2013),
General circulation modeling of the Martian upper atmosphere during global
dust storms, \textit{J. Geophys. Res.}, \textit{118}, 2234--2246,
doi:10.1002/2013JE004429.

\bibitem[{\textit{Medvedev et al.}(2011a)}]{myh11}
Medvedev,~A.~S., E.~Yi\u{g}it, and P.~Hartogh (2011a), Estimates of gravity
wave drag on Mars: Indication of a possible lower thermospheric wind
reversal, \textit{Icarus}, \textit{211}, doi:10.1016/j.icarus2010.10.013,
909--912.

\bibitem[{\textit{Medvedev et al.}(2011b)}]{myhb11}
Medvedev,~A.~S., E.~Yi\u{g}it, P.~Hartogh, and E. Becker (2011b),
Influence of gravity waves on the Martian atmosphere: general circulation
modeling, \textit{J. Geophys. Res.}, \textit{116}, E10004,
doi:10.1029/2011JE003848.

\bibitem[{\textit{Medvedev and Hartogh}(2007)}]{MH07}
Medvedev, A.~S., and P. Hartogh (2007), Winter polar warmings and the
meridional transport on Mars simulated with a general circulation model,
\textit{Icarus}, \textit{186}, 97--110.

\bibitem[{\textit{Medvedev and Klaassen}(2003)}]{MK03}
Medvedev,~A.~S., and G.~P. Klaassen (2003),
Thermal effects of saturating gravity waves in the atmosphere,
\textit{J. Geophys. Res.}, \textit{108}, 4040, doi:10.1029/2002JD002504.

\bibitem[{\textit{Montmessin et al.}(2004)}]{mont04}
Montmessin, F., F. Forget, P. Rannou, M. Cabane, and R. M. Haberle (2004),
Origin and role of water ice clouds in the Martian water cycle as inferred
from a general circulation model, \textit{J. Geophys. Res.}, \textit{109},
E10004, doi:10.1029/2004JE002284.

\bibitem[{\textit{Nair et al.}(1994)}]{Nair94}
Nair, H., M. Allen, A.~D. Anbar, Y.~L. Yung, and R.~T. Clancy (1994),
A photochemical model of the martian atmosphere, \textit{Icarus},
\textit{111}, 124--150.

\bibitem[{\textit{Nakajima et al.}(2000)}]{nakaj00}
Nakajima, T., M. Tsukamoto, Y. Tsushima, A.~Numaguti, and T.~Kimura
(2000), Modeling of the radiative processes in an atmospheric general
circulation model, \textit{Applied Optics}, \textit{39}(N27), 4869--4878.

\bibitem[{\textit{Nakajima and Tanaka}(1986)}]{nakaj86}
Nakajima, T., and M. Tanaka (1986), Matrix formulations for the transfer of
solar radiation in a plane-parallel scattering atmosphere,
\textit{J. Quant.  Spectrosc. Radiat. Transfer}, \textit{35}, 13--21.

\bibitem[{\textit{Qu\'{e}merais et al.}(2006)}]{quem06}
Qu\'{e}merais, E., J.-L. Bertaux, O. Korablev, E. Dimarellis, C. Cot,
B.~R. Sandel, and D. Fussen (2006), Stellar occultations observed by SPICAM
on Mars Express, \textit{J. Geophys. Res.}, \textit{111}, E09S04,
doi:10.1029/2005JE002604.

\bibitem[{\textit{Richards et al.}(1994)}]{richards94}
Richards, P.~G., J. A. Fennely, and D.~G. Torr (1994), EUVAC: A solar EUV
flux model for aeronomic calculations, \textit{J. Geophys. Res.},
\textit{99}, 8981--8992.

\bibitem[{\textit{Stewart et al.}(1992)}]{stewart92}
Stewart, A.~I.~F., M. J. Alexander, R.~R. Meier, L.~J. Paxton, S.~W. Bougher,
and C.~G. Fesen (1992), Atomic oxygen in the Martian thermosphere,
\textit{J. Geophys. Res.}, \textit{97}(A1), 91--102.

\bibitem[{\textit{Tolson et al.}(2002)}]{tolson02}
Tolson, R.~H., G.~M. Keating, B.~E. George, P.~E. Escalera, M.~R. Werner,
A.~M. Dwyer, and J.~L. Hanna (2002), Application of accelerometer data
to Mars Odyssey aerobraking and atmospheric modeling,
\textit{J. Spacecraft Rockets}, \textit{42}, 435--443.

\bibitem[{\textit{Tolson et al.}(2007)}]{tolson07}
Tolson, R., G. Keating, R.~W. Zurek, S. W. Bougher, C.~J. Justus,
D.~C. Fritts (2007), Application of accelerometer data to atmospheric
modeling during Mars aerobraking operations, \textit{J. Spacecraft
Rockets}, \textit{44}, 1172--1179.

\bibitem[{\textit{Valeillet et al.}(2009)}]{vale09}
Valeille, A., V. Tenishev, S.W. Bougher, M.R. Combi, and A. Nagy (2009),
Three-dimensional study of Mars upper thermosphere/ionosphere and hot oxygen
coronae: 1. General description and results at equinox for low solar
conditions, \textit{J. Geophys. Res.}, \textit{114}, E11005,
doi:1029/2009JE003388.

\bibitem[{\textit{Withers and Pratt}(2013)}]{withers13}
Withers, P., and R. Pratt (2013), An observational study of the response of
the upper atmosphere of Mars to lower atmospheric dust storms, \textit{Icarus},
\textit{225}, 378--389.

\bibitem[{\textit{Withers}(2006)}]{withers06}
Withers, P.G. (2006), MGS and Mars Odyssey accelerometer observations of the
martian upper atmosphere during aerobraking, \textit{Geophys. Res. Lett.},
\textit{33}, L02201, doi:10.1029/2005GL024447.

\bibitem[{\textit{Yagi et al.}(2009)}]{yagi12}
Yagi, M., F. Leblanc, J. Y. Chaufray, F. Gonzalez-Galindo, S. Hess,
and R. Modolo (2012), Mars exospheric thermal and non-thermal components:
Seasonal and local variations, \textit{Icarus}, \textit{221}, 682--693,
doi:10.1016/j.icarus.2012.07.022.

\bibitem[{\textit{Yi\u{g}it and Medvedev}(2009)}]{YigitMedvedev09}
  {Yi\u{g}it, E., and A.~S. Medvedev} (2009), Heating and cooling of the
  thermosphere by internal gravity waves, \textit{Geophys. Res. Lett.},
  \textit{36}, L14807, \doi{10.1029/2009GL038507}.

\bibitem[{\textit{Yi\u{g}it and Medvedev}(2010)}]{YM10}
Yi\u{g}it,~E., and A.~S.~Medvedev (2010), Internal gravity waves in the
thermosphere during low and high solar activity: Simulation study,
\textit{J. Geophys. Res.}, \textit{115}, A00G02, doi:10.1029/2009JA015106.

\bibitem[{\textit{Yi\u{g}it and Medvedev}(2012)}]{YigitMedvedev12}
Yi\u{g}it, E., and A.~S. Medvedev (2012), Gravity waves in the
  thermosphere during a sudden stratospheric warming, \textit{Geophys. Res.
  Lett.}, \textit{39}, L21101, doi:10.1029/2012GL053812.

\bibitem[{\textit{Yi\u{g}it and Medvedev}(2015)}]{YigitMedvedev15}
Yi\u{g}it, E., and A.~S. Medvedev (2015), Internal wave coupling processes in
  Earth's atmosphere, \textit{Adv. Space Res., 55}, 983--1003,
  \doi{10.1016/j.asr.2014.11.020}.

\bibitem[{\textit{Yi\u{g}it et~al.}(2008)}]{Yigit_etal08}
Yi\u{g}it,~E., A.~D.~Aylward, and A.~S.~Medvedev (2008), Parameterization of
the effects of vertically propagating gravity waves for thermosphere general
circulation models: Sensitivity study, \textit{J. Geophys. Res.}, \textit{113},
D19106, doi:10.1029/2008JD010135.

\bibitem[{\textit{Yi\u{g}it et~al.}(2009)}]{Yigit_etal09}
Yi\u{g}it, E., A.~S. Medvedev, A.~D. Aylward, P.~Hartogh, and M.~J.
  Harris (2009), Modeling the effects of gravity wave momentum deposition on
  the general circulation above the turbopause, \textit{J. Geophys. Res.},
  \textit{114}, D07101, \doi{10.1029/2008JD011132}.

\bibitem[{\textit{Yi\u{g}it et~al.}(2012)}]{Yigit_etal12}
Yi\u{g}it, E., A.~S. Medvedev, A.~D. Aylward, A.~J. Ridley, M.~J.
  Harris, M.~B. Moldwin, and P.~Hartogh (2012), Dynamical
  effects of internal gravity waves in the equinoctial thermosphere,
  \textit{J. Atmos. Sol.-Terr. Phys.}, \textit{90--91}, 104--116,
  \doi{10.1016/j.jastp.2011.11.014}.

\bibitem[{\textit{Yi\u{g}it et~al.}(2014)}]{Yigit_etal14}
Yi\u{g}it, E., A.~S. Medvedev, S.~L. England, and T.~J. Immel (2014),
  Simulated variability of the high-latitude thermosphere induced by
  small-scale gravity waves during a sudden stratospheric warming, \textit{J.
  Geophys. Res. Space Physics}, \textit{119}, \doi{10.1002/2013JA019283}.

\end{thebibliography}
\end{document}